\documentclass[aps,prl,reprint]{revtex4-2}
\usepackage{graphicx}
\usepackage{natbib}
\usepackage{amssymb}
\usepackage{amsmath}
\usepackage[utf8]{inputenc} 
\usepackage{braket}
\usepackage{graphicx}
\usepackage{subfigure}
\usepackage{pgfplots}
\usepackage{csquotes}
\usepackage{hhline}
\usepackage{amssymb}

\usepackage{dcolumn}
\usepackage{tabularx}
\usepackage[colorlinks=true,linkcolor=blue,citecolor=blue,urlcolor=blue]{hyperref}
\usepackage{longtable}
\usepackage{braket}
\usepackage{float}
\usepackage{listings}
\usepackage{color}
\definecolor{mygreen}{rgb}{0,0.6,0}
\definecolor{mygray}{rgb}{0.5,0.5,0.5}
\definecolor{mymauve}{rgb}{0.58,0,0.82}

\newcolumntype{C}{>{\centering\arraybackslash}X}

\begin{document}

\title{Localization, $\mathcal{PT}$-Symmetry Breaking and Topological Transitions in non-Hermitian Quasicrystals}
\author{Aruna Prasad Acharya}
\affiliation{Department of Physics and Astronomy, National Institute of Technology, Rourkela, Odisha-769008, India}
\author{Aditi Chakrabarty}
\affiliation{Department of Physics and Astronomy, National Institute of Technology, Rourkela, Odisha-769008, India}
\author{Deepak Kumar Sahu}
\affiliation{Department of Physics and Astronomy, National Institute of Technology, Rourkela, Odisha-769008, India}
\author{Sanjoy Datta}
\email{dattas@nitrkl.ac.in}
\affiliation{Department of Physics and Astronomy, National Institute of Technology, Rourkela, Odisha-769008, India}
\affiliation{Center for Nanomaterials, National Institute of Technology, Rourkela, Odisha-769008, India}
\date{\today}

\begin{abstract}
According to the topological band theory of a Hermitian system, the different electronic phases are 
classified in terms of topological invariants, wherein the transition between the two phases characterized 
by a different topological invariant is the primary signature of a topological phase transition.
Recently, it has been argued that the delocalization-localization transition in a quasicrystal, described by
the non-Hermitian $\mathcal{PT}$-symmetric extension of the Aubry-Andr\'{e}-Harper (AAH) Hamiltonian
can also be identified as a topological phase transition. Interestingly, the $\mathcal{PT}$-symmetry
also breaks down at the same critical point. However, in this article, we have shown that the 
delocalization-localization transition and the $\mathcal{PT}$-symmetry breaking are not connected 
to a topological phase transition. To demonstrate this, we have studied the non-Hermitian $\mathcal{PT}$-symmetric 
AAH Hamiltonian 
in the presence of Rashba Spin-Orbit (RSO) coupling. We have obtained an analytical expression of 
the topological transition point and compared it with the numerically obtained critical points. We have 
found that, except in some special cases, the critical point and the topological transition point are not 
the same. In fact, the delocalization-localization transition takes place earlier than the topological
transition whenever they do not coincide.
\end{abstract}

\maketitle
\indent\emph{Introduction}- 
The discovery of the connection between the electronic phases of a Hermitian system and the topology of its 
band structure has had far reaching consequences in a wide range of electronic phenomena, such as 
topological insulators \cite{Hasan,Lau,Qi}, semimetals \cite{Armitage,Ganeshan2015} and superconductors 
\cite{Abanov}. In this approach, an electronic phase is characterized by a topological quantity which remains
invariant under the continuous deformation of some control parameter in the Hamiltonian. A topological transition
takes place whenever there is a change in the electronic phases, characterized by a different topologically 
invariant quantity, as the control parameter changes \cite{C_liu,Kraus2012,Lang,Bansil}. 
Naturally, the question arises regarding the existence of such topological transitions in a $\mathcal{PT}$-symmetric 
non-Hermitian system. In the last decade, several investigations on the topological phenomena in the
non-Hermitian $\mathcal{PT}$-symmetric systems have been realized \cite{Ni,Klett,Zhu,Pablo}.

Recently, it has been pointed out that the delocalization-localization transition in a 
non-Hermitian $\mathcal{PT}$-symmetric extension of the Aubry-Andr\'{e}-Harper (AAH) Hamiltonian
can also be classified as a topological transition between the two topologically non-trivial 
phases characterized by a different topological invariant \cite{Longhi}. Moreover, the $\mathcal{PT}$-symmetry also 
breaks down at this critical point. Several instances of the topological transitions being accompanied with the 
localization-delocalization transition \cite{Tang} 
and the existence of mobility edge \cite{Zeng} have been investigated for years. 
In a recent work \cite{XCai}, by introducing an additional non-hermiticity in the hopping parameters, 
it has been reported that the delocalization-localization is not necessarily in accordance 
with the topological phase transitions.

The objective of this Letter is to demonstrate that the localization-delocalization transition and the $\mathcal{PT}$-symmetry
breaking are not related to a topological transition in the $\mathcal{PT}$-symmetric non-Hermitian quasicrystals which 
does not require any additional breaking of the hermiticity.  
To demonstrate this concretely, we have considered the non-Hermitian AAH Hamiltonian 
in the presence of the Rashba Spin-Orbit (RSO) coupling, conserving the $\mathcal{PT}$-symmetry of the system. As a first step, we obtain an analytical expression of the 
topological transition point. The critical point, which is also identical with the $\mathcal{PT}$-symmetry breaking point,
is determined numerically. We have found that the critical point does not coincide with the analytically obtained 
topological transition point, except in some special cases.  \\
\begin{figure*}[ht]
	\centering 
	\includegraphics[width=0.317\textwidth,height=0.26\textwidth]{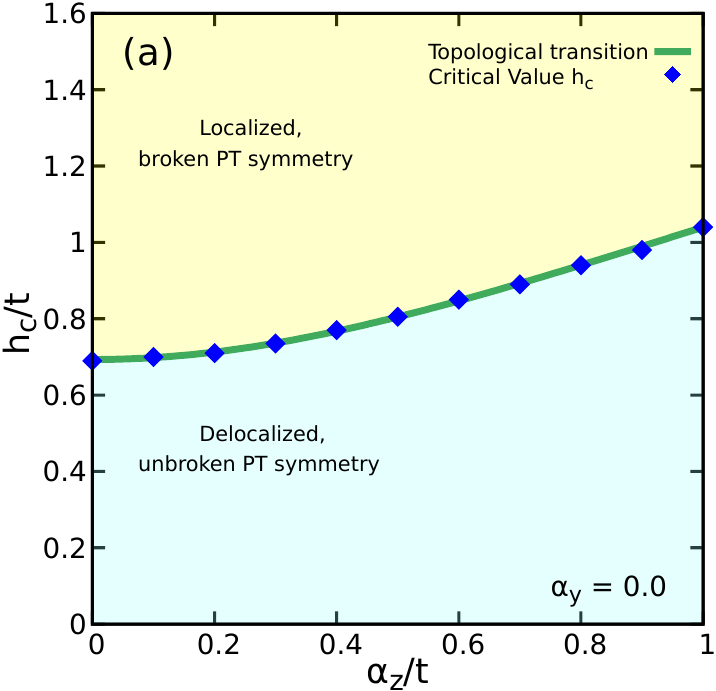}\hspace{0.2cm}
	\includegraphics[width=0.32\textwidth,height=0.26\textwidth]{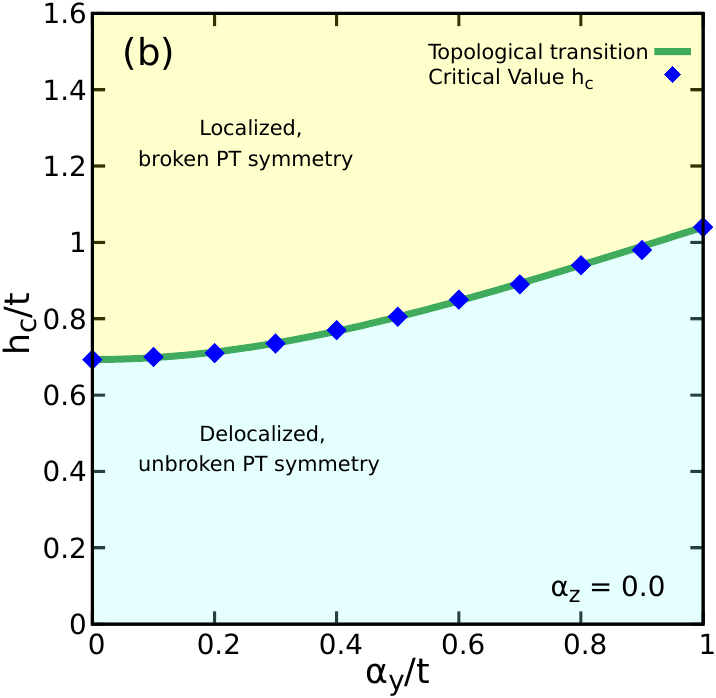}\hspace{0.1cm} 
	\includegraphics[width=0.32\textwidth,height=0.26\textwidth]{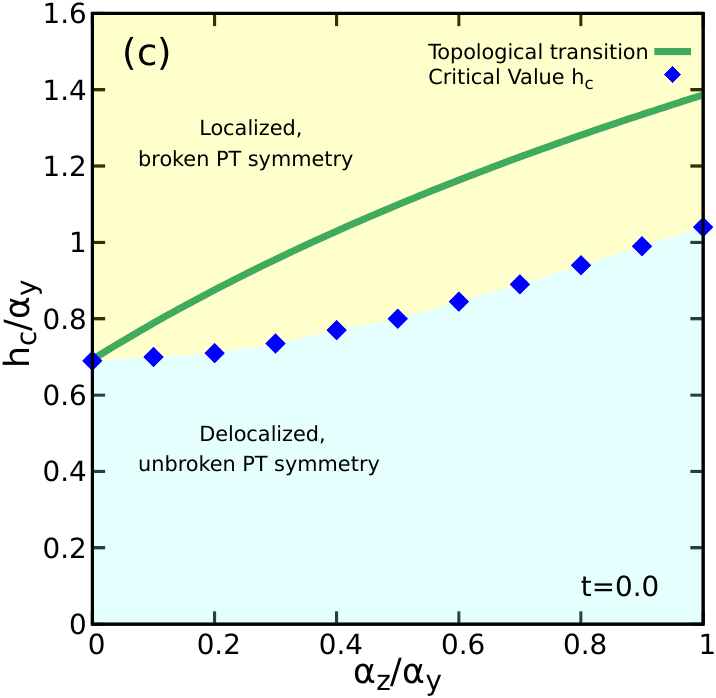}\hspace{0.1cm}\\
	\includegraphics[width=0.32\textwidth,height=0.26\textwidth]{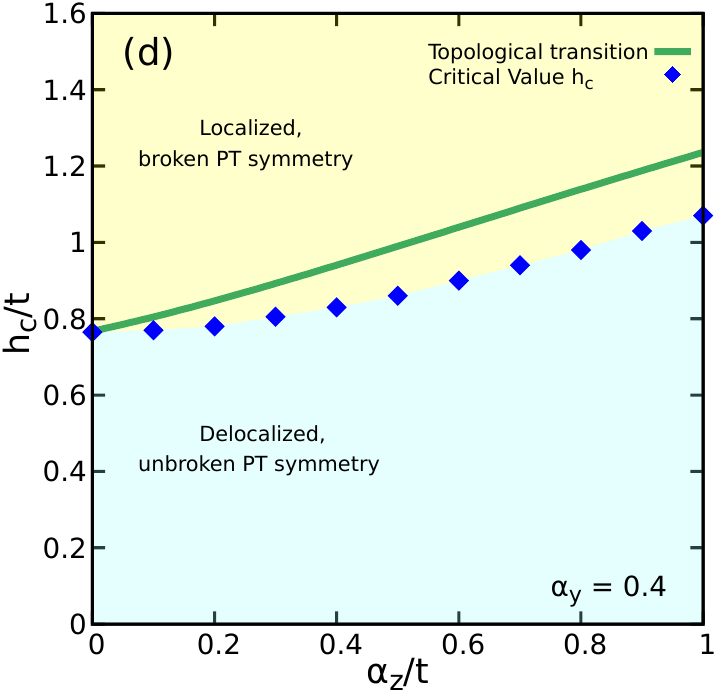} \vspace{-0.2cm}\hspace{0.1cm}
	\includegraphics[width=0.32\textwidth,height=0.26\textwidth]{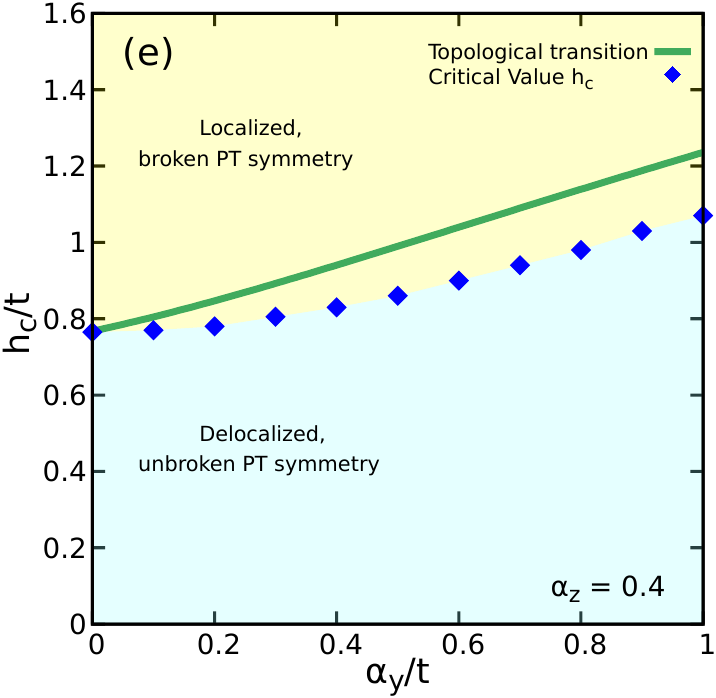}\hspace{0.1cm}
	\includegraphics[width=0.32\textwidth,height=0.26\textwidth]{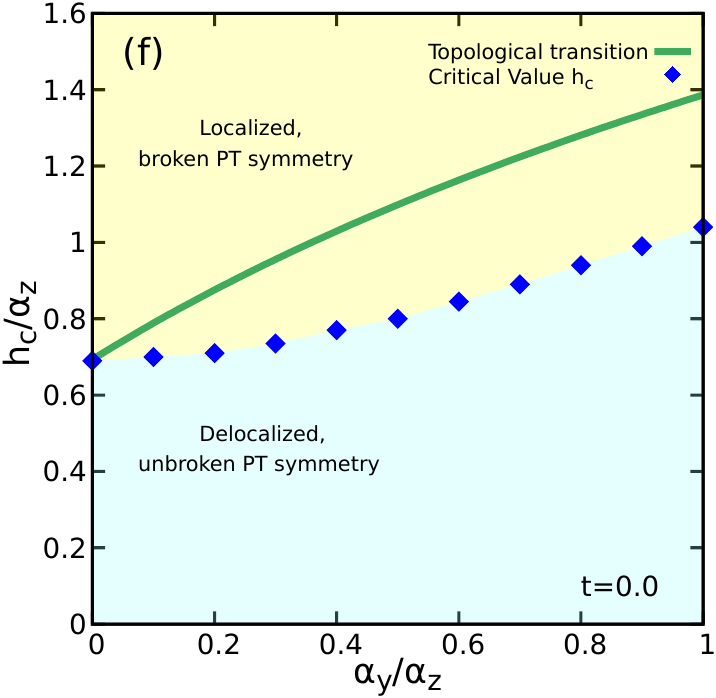} \hspace{0.1cm}
	\vspace{-0.2cm}
	\caption{The variation of the critical value $h_c$ with the strength of RSO coupling. The numerically calculated values of $h_c$
	have been shown by diamond shaped markers (in blue), and the analytical result of $h_t$ from Eq. (\ref{Eq:RSO_Lyapunov_Exponent}) has been represented by solid lines (in green). The results are for a lattice with 610 sites, and obeying the periodic boundary conditions. 
	(a) and (b): Phase diagram of the critical points when one of the RSO hopping amplitude ($\alpha_y$ or $\alpha_z$) is 
	set to zero, while the other component varies with a non-zero amplitude. (d) and (e): Phase diagram of the critical points 
	when $\alpha_y/t$ is set to 0.40 while the component $\alpha_z$ varies, and vice-versa. 
	(c) and (f): Phase diagram of the critical points with $t$ set to zero, and when $\alpha_y$ varies as a function of 
	$\alpha_z$ and vice-versa.} 
	\label{Fig: Phase diagram for h_c}
\end{figure*}
\indent \emph{Model-} The Hamiltonian considered in this Letter originates from the model proposed by Longhi 
\cite{Longhi}, comprising non-interacting electrons in an one-dimensional quasiperiodic lattice with a $\mathcal{PT}$ symmetric 
non-Hermiticity introduced in the cosine-modulated phase of the conventional AAH model  
\cite{Harper, Aubry} and is given by,
\begin{eqnarray}
H_A &=& -t\displaystyle\sum_{{n=1},\sigma}^{L-1} (c^\dag_{{n+1},{\sigma}} c_{{n},{\sigma}}+ h.c) + 
\sum_{{n=1},{\sigma}}^{L} V_n c^\dag_{{n},{\sigma}} c_{{n}{\sigma}},~~~~ \label{Eq:PT-AAH-Hamiltonian}
\end{eqnarray}
where $t$ is the amplitude of hopping integral in the tight-binding Hamiltonian, and $n$ is the lattice site 
index in the real space and $L$ denotes the size of the lattice. $c^\dag_{{n},{\sigma}}$ and $c_{{n},{\sigma}}$ represent the electronic creation and annihilation operators respectively at site \emph{n} with spin state $\sigma$ ($\sigma=\uparrow$,$\downarrow$).
The non-Hermiticity is introduced in the form of a complex phase in the on-site potential and is expressed as,
\begin{equation}
   V_n=V \rm{cos}(2\pi{\alpha}n + \phi),
\end{equation}
where $\phi=\theta+ih$. $V$ is the amplitude of the modulation, and $\alpha$ is an irrational number from the Diophantine approximation
($\alpha=(\sqrt{5}+1)/2$). 
When $h=0$, we get back the original Hermitian version of the AAH Hamiltonian. 
It is easy to see that $H_A$ retains the $\mathcal{PT}$-symmetry after the introduction of this complex phase. In the presence of the RSO coupling, our Hamiltonian $H$ consists of two parts, and is given by,
\begin{equation}
\centering{H=H_A+H_R},
\label{Eq:Full_Hamiltonian}
\end{equation}
where $H_A$ is the non-Hermitian counterpart of the standard AAH Hamiltonian given by Eq. (\ref{Eq:PT-AAH-Hamiltonian}), and  
$H_R$ includes the spin conserving hopping amplitude$(\alpha_y)$ and the spin-flip hopping amplitude$(\alpha_z)$ induced 
by RSO coupling and can be written as \cite{Birkholz,Mireles},
\begin{eqnarray} 
H_R &=& -\alpha_z \displaystyle\sum_{{n=1},{\sigma},{\sigma'}}^{L-1} (c^\dag_{n+1,\sigma}(i\sigma_y)_{\sigma,\sigma'}
 c_{n,\sigma'}+ h.c.) \nonumber \\
 & & -\alpha_y \displaystyle\sum_{{n=1},{\sigma},{\sigma'}}^{L-1} (c^\dag_{n+1,\sigma}(i\sigma_z)_{\sigma,\sigma'}  
 c_{n,\sigma'}+ h.c.).
\end{eqnarray}\\ 
\begin{figure*}[ht]
	\centering
	\includegraphics[width=0.242\textwidth,height=0.223\textwidth]{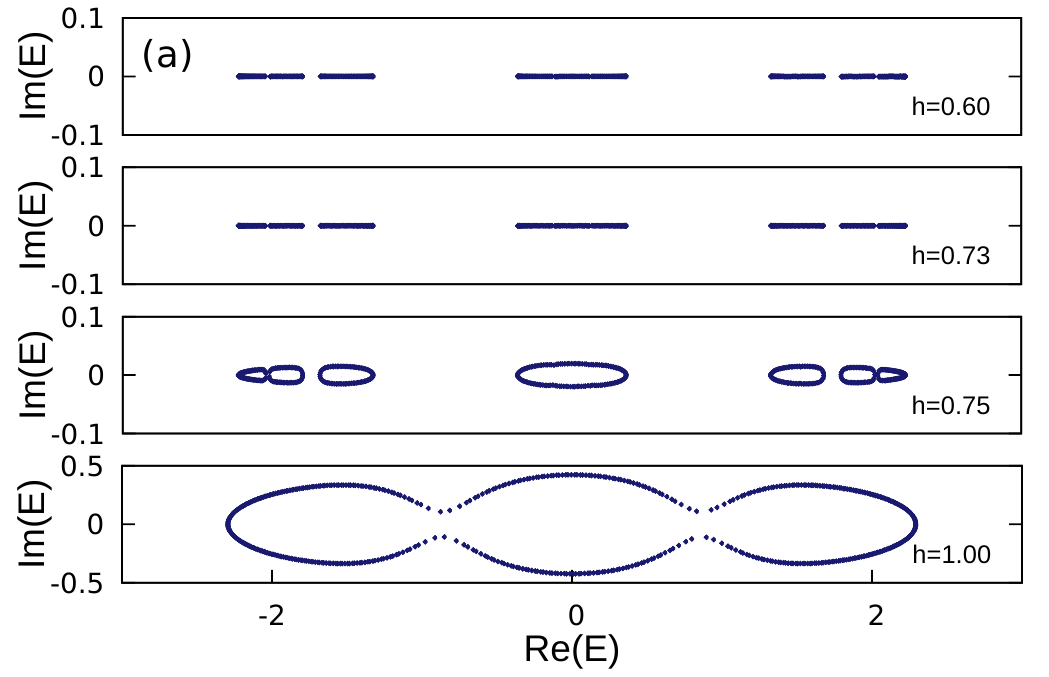}\hspace{0.15cm}
	\includegraphics[width=0.24\textwidth,height=0.22\textwidth]{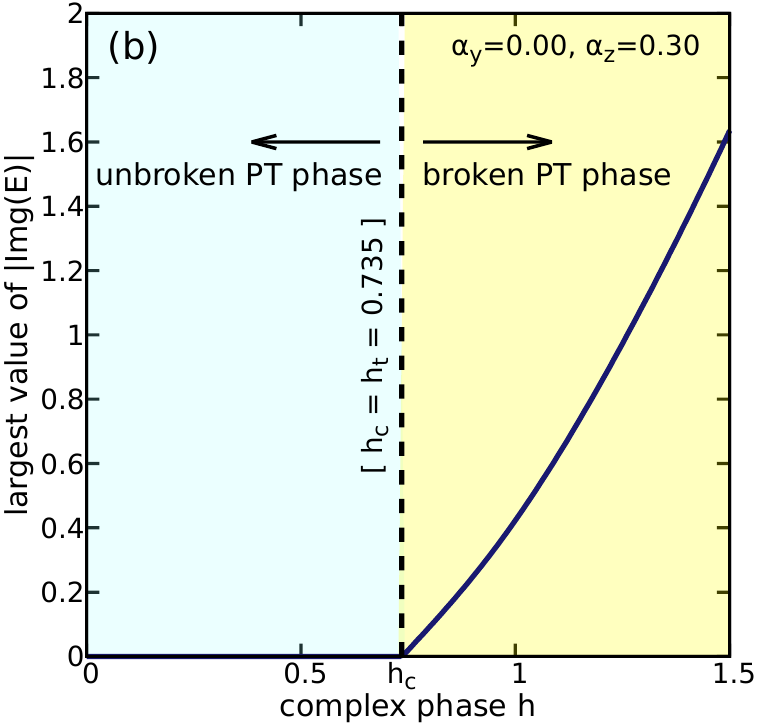}\hspace{0.06cm}
	\includegraphics[width=0.242\textwidth,height=0.22\textwidth]{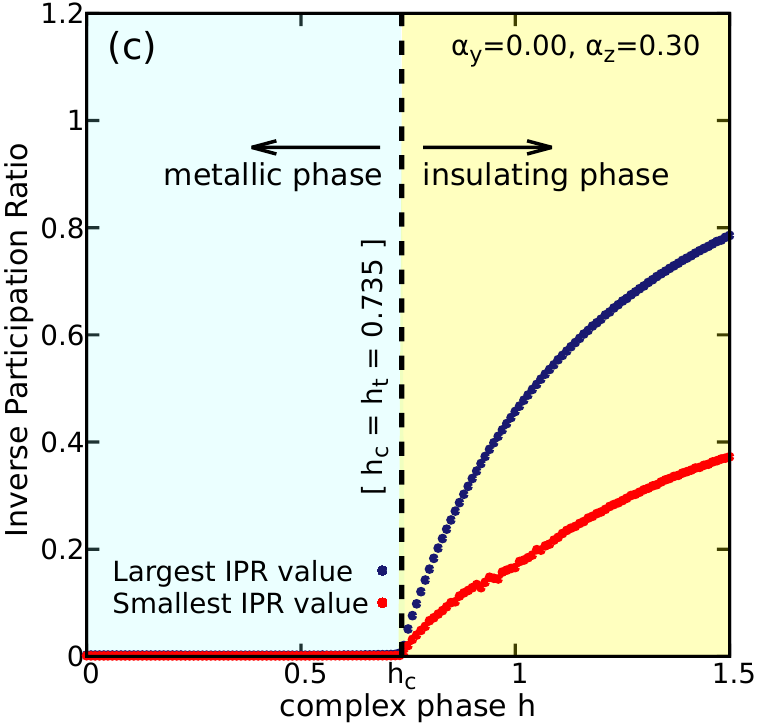}\hspace{0.1cm}
	\includegraphics[width=0.24\textwidth,height=0.22\textwidth]{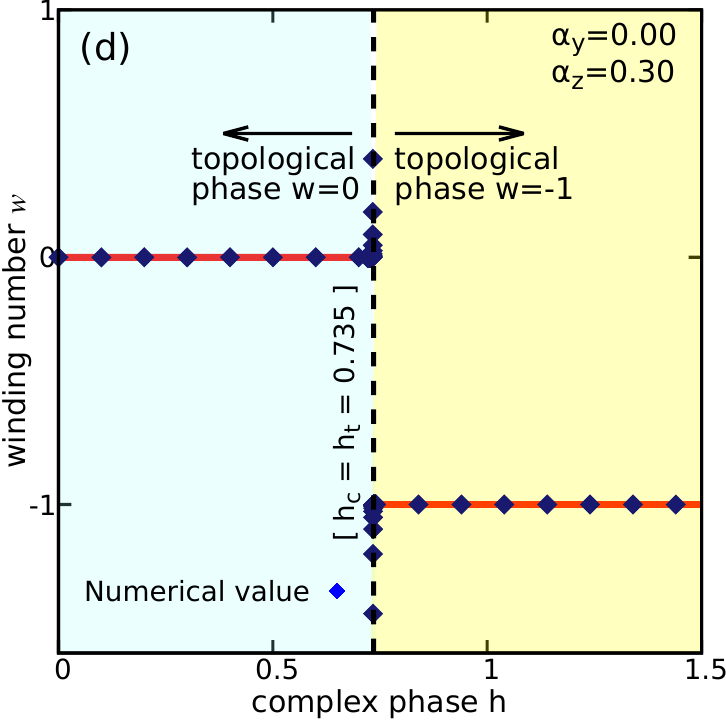}  \vspace{0.03 cm}\\
	\includegraphics[width=0.243\textwidth,height=0.224\textwidth]{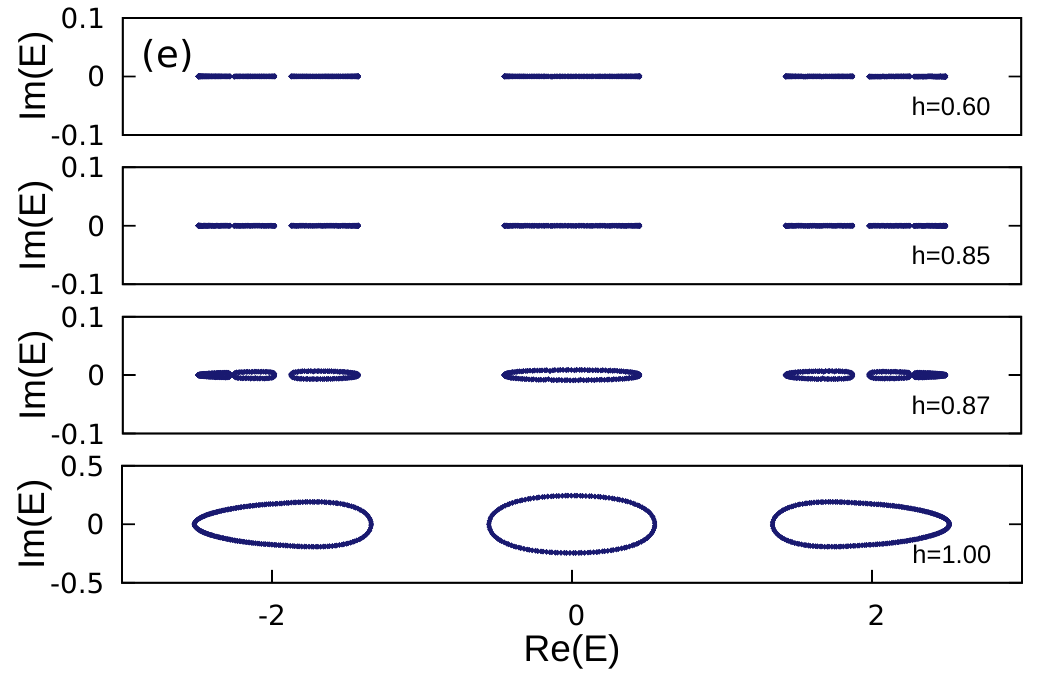}\hspace{0.12cm}
	\includegraphics[width=0.24\textwidth,height=0.222\textwidth]{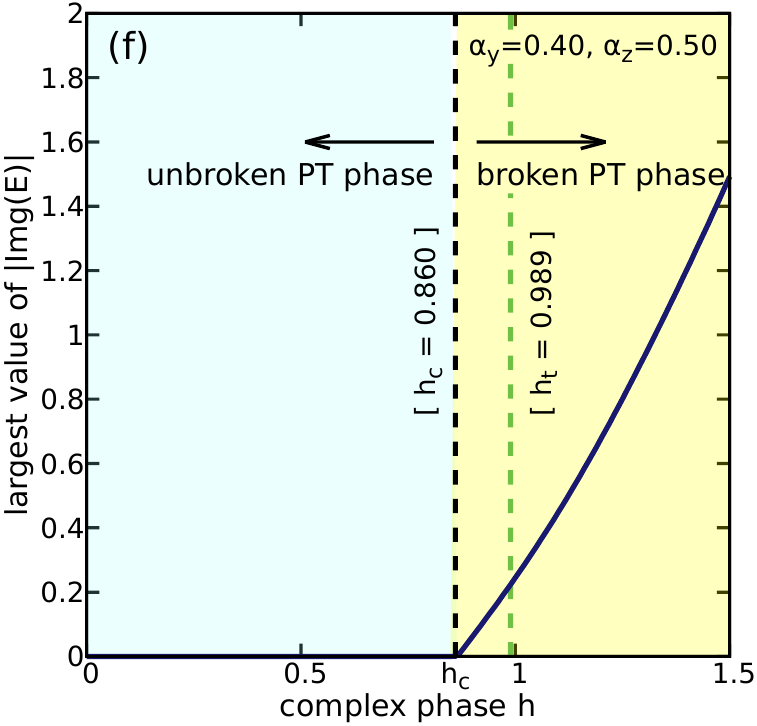}\hspace{0.12cm}
	\includegraphics[width=0.238\textwidth,height=0.222\textwidth]{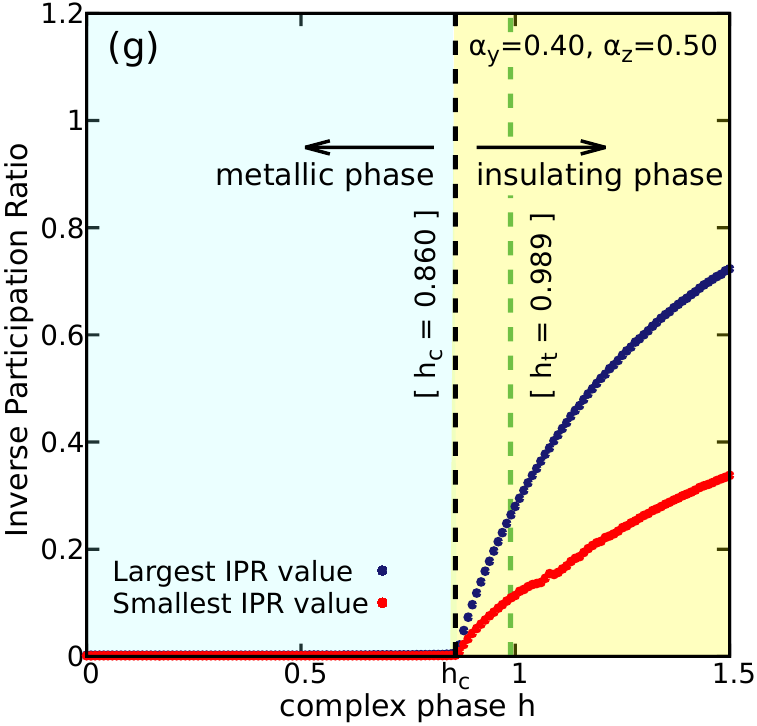}\hspace{0.14cm}
	\includegraphics[width=0.236\textwidth,height=0.221\textwidth]{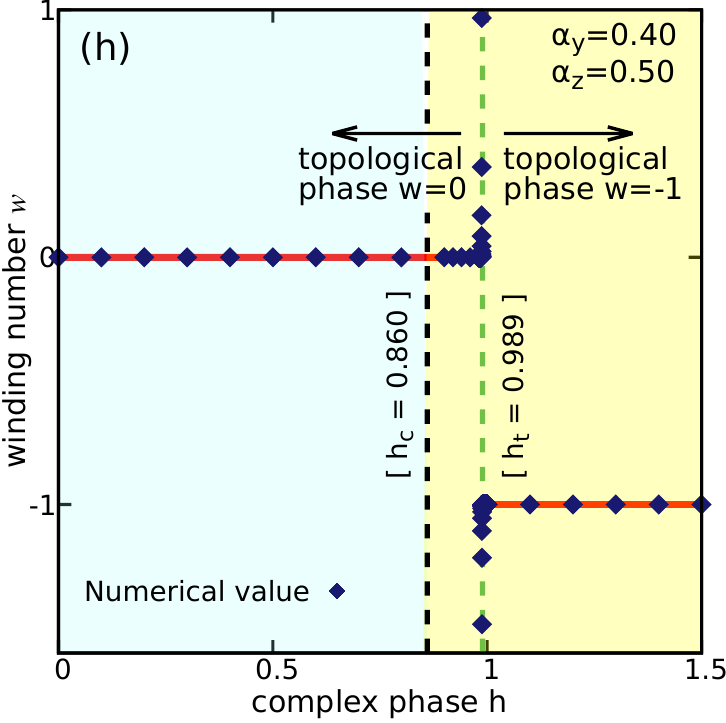}
	\caption{Phase transition in non-Hermitian quasicrystals described by Eq. (\ref{Eq:quasi-AAH-eigen-equivalent}), 
	where $t$ has been set equal to $V$ ($\emph{t=V=1}$), and 
	with $\theta=0$. The upper panel is for the RSO coupling strength set at $\alpha_y=0$ 
	and $\alpha_z=0.3$, while in the lower panel $\alpha_y=0.4$ and $\alpha_z=0.5$. (a) and (e): The energy spectrum 
	spanned in the complex plane, with an increasing value of the control parameter $\emph{h}$. (b) and (f): The 
	nature of the largest value of $|Im(E)|$ $\emph{vs. h}$ across the PT symmetry breaking point. The horizontal 
	black dotted line depicts the critical value $h_c$, whereas the green dotted line shows the transition point $h_t$ determined analytically in Eq. (\ref {Eq:RSO_Lyapunov_Exponent}). (c) and (g): Behavior of the largest(in blue)
	and smallest(in red) values of the Inverse Participation Ratio with an increasing value of $\emph{h}$, before and after 
	the metal-insulator transition. (d) and (h): The variation of numerically determined values of winding number $\emph{w(h)}$ with $h$(in blue) and the analytical result from Eq. (\ref{Eq:winding_number_final}) shown as solid red line.}
	\label{Fig:Phase transition and winding number}
\end{figure*}
\indent 
To find the analytical expression of the topological transition point in the presence of RSO coupling, at first, we start with 
the eigenvalue equation $H \left| \Psi \right\rangle = E \left| \Psi \right\rangle$, where $\Psi$ is a linear superposition of 
the spin states given by, $\left| \Psi \right\rangle = \sum_{n,\sigma} \psi_{n}^{\sigma} c^{\dagger}_{n,\sigma} \left|0 \right\rangle$. 
Taking into account the two spin orientations, the set of coupled eigenvalue equations read as,
\begin{eqnarray}
-t\left(\psi_{n+1}^{\uparrow} + \psi_{n-1}^{\uparrow} \right) 
+\alpha_y\left(e^{-i \pi/2}\psi_{n+1}^{\uparrow} +  e^{i \pi/2}\psi_{n-1}^{\uparrow} \right) & &\nonumber \\
+\alpha_z \left(\psi_{n+1}^{\downarrow} - \psi_{n-1}^{\downarrow} \right)
+ V~\rm{cos}(2\pi\alpha n + \phi) \psi_{n}^{\uparrow} = E \psi_{n}^{\uparrow},~~~~~~~~\label{Eq:eigen-eq-sup}\\
-t\left(\psi_{n+1}^{\downarrow} + \psi_{n-1}^{\downarrow} \right) 
+ \alpha_y\left(e^{-i \pi/2} \psi_{n+1}^{\downarrow} + e^{i \pi/2}\psi_{n-1}^{\downarrow} \right) \nonumber \\
-\alpha_z \left(\psi_{n+1}^{\uparrow} - \psi_{n-1}^{\uparrow} \right)
+ V~\rm{cos}(2\pi\alpha n + \phi) \psi_{n}^{\downarrow} = E \psi_{n}^{\downarrow}.~~~~~~~~\label{Eq:eigen-eq-sdn}
\end{eqnarray}
Multiplying $i$ to Eq.~(\ref{Eq:eigen-eq-sdn}) and adding it to Eq.~(\ref{Eq:eigen-eq-sup}), we obtain a single 
eigenvalue equation given by,
\begin{eqnarray}
t'\left[  e^{-i \eta} \tilde\psi_{n+1} + \tilde\psi_{n-1} e^{ i \eta} \right] 
+ V \rm{cos}(2\pi\alpha n + \phi) \tilde\psi_{n}
=E \tilde\psi_{n}, ~~~~\nonumber \label{Eq:eigen-eq-quasiparticle}\\
\end{eqnarray}
where $\tilde\psi_{n}=\psi_{n}^{\uparrow}+i\psi_{n}^{\downarrow}$, $t'=\sqrt{t^2 +(\alpha_y+\alpha_z)^2}$ and 
$\eta=\rm{tan}^{-1}\left[-(\alpha_y +\alpha_z)/t\right]$.\\

In the next step, we are going to show that the above eigenvalue equation is identical with the 
non-Hermitian $\mathcal{PT}$-symmetric AAH Hamiltonian expressed in Eq.~(\ref{Eq:PT-AAH-Hamiltonian}). To do 
this, we obtain the eigenvalue equation in momentum space with the help of the Fourier transformation 
$\tilde \psi_{n} = \sum_{k} \tilde \psi_{k} e^{i \left(2 \pi\alpha n\right)k}$ to get:
\begin{eqnarray}
W_k  \tilde\psi_{k}+ \frac{V}2 (e^{i \phi} \tilde \psi_{k+1}+ \tilde \psi_{k-1}  e^{-i \phi})=E \tilde \psi_{k}, 
\label{Eq:quasi-Fourier-eigenvalue}
\end{eqnarray}
where,
\begin{eqnarray}
W_k=2 t' \rm{cos}(2\pi\alpha k - \eta),
\end{eqnarray}
and $\tilde\psi_{k} = \psi_{k}^{\uparrow} + i \psi_{k}^{\downarrow}$. 
In the final step, we apply an inverse Fourier transformation 
$\tilde \psi_{k} = \sum_{n}\tilde \psi_{n} e^{-i \left(2 \pi \alpha k - \eta \right)n}$ on
Eq.~(\ref{Eq:quasi-Fourier-eigenvalue}) to get:
\begin{eqnarray}
t'\left[ \tilde\psi_{n+1} + \tilde\psi_{n-1} \right] + V \rm{cos}(2\pi \alpha n+\phi) \tilde\psi_{n}
=E \tilde\psi_{n}. \label{Eq:quasi-AAH-eigen-equivalent}
\end{eqnarray}
It is quite evident that the eigenvalue equation in Eq.~(\ref{Eq:quasi-AAH-eigen-equivalent}) is identical with 
that of Eq.~(\ref{Eq:PT-AAH-Hamiltonian}), with the rescaled hopping amplitude $t'$ in lieu of the tight 
binding hopping amplitude $t$.

The common approach to identify the critical point in this type of quasicrystals is to find the  
Lyapunov exponent characterizing the inverse of the localization length \cite{Aubry,Sokoloff}. In Ref.~\cite{Longhi} it 
has been established that this approach is equally applicable for the original $\mathcal{PT}$-symmetric non-Hermitian 
AAH Hamiltonian without the RSO coupling. In this case, the Lyapunov exponent is given by  $\gamma=\log(2t/V)$ and there is a 
phase transition from an extended/metallic ($V<2t$) to a localized/insulating regime ($V>2t$).  
Such a transition also coincides with the PT symmetry breaking point, at which the energy spectrum changes from 
real(and positive) to entirely complex \cite{Bender,Bender2007}. Since, in the presence of the RSO coupling, our 
Hamiltonian is equivalent to the original $\mathcal{PT}$-symmetric AAH Hamiltonian, it is anticipated that
in the presence of the RSO coupling the $\mathcal{PT}$-symmetric non-Hermitian AAH Hamiltonian would undergo
a similar delocalization-localization transition at $V=2t'$. In this case, the Lyapunov exponent is given by, 
$\gamma=\log(2t'/V)$. However, we are going to show that this is \emph{not true} in general. We have found that when
either all the hopping amplitudes are non-zero or the tight-binding hopping amplitude is zero, 
the localization-delocalization transition and the $\mathcal{PT}$-symmetry breaking still coincide with 
each other, but takes place earlier than the anticipated value from the analysis of the Lyapunov exponent.
Interestingly, however, we have found that the anticipated value of the transition point is associated with 
a topological transition. This \emph{clearly} indicates that the delocalization-localization transition in 
the non-Hermitian $\mathcal{PT}$-symmetric AAH Hamiltonian is not connected to a topological transition. 

\indent \emph{Behavior of the topological phases, symmetry breaking, metal-insulator transition and phase diagram in 
the presence of RSO}- 
\indent 
The method to classify the phases based on topology of the band structure is quite well established by now 
for the Hermitian systems \cite{Hasan,Bansil}. Recently, Gong \emph{et al.} \cite{Gong} have 
laid down a method for the topological classification of the electronic phases in non-Hermitian systems. Similar to the spirit for classifying phases in an Hermitian system, the above-mentioned method has been used in Ref.~\cite{Longhi} to define a winding number, separating the two topological phases.
Here, following the same approach laid down by Gong \emph{et al.} \cite{Gong} and Longhi \cite{Longhi}, 
we assume the complex phase \emph{h} as a control parameter of the Hamiltonian, where $H(\phi)=H(\theta,h)$
is given by,
\begin{eqnarray}
H=
\left( 
\begin{matrix}
V_1 & t' & 0 & 0 & 0 & ... & 0 & 0 & t'  \\
t' & V_2 & t' & 0 & 0 & ... & 0 & 0 & 0 \\
0 & t' & V_3 & t' & 0 & ... & 0 & 0 & 0 \\
... & ... & ... & ... & ... & ... & ... & ... & ... \\
0 & 0 & 0 & 0 & 0 & ... & t' & V_{L-1} & t' \\
t' & 0 & 0 & 0 & 0 & ... & 0 & t' & V_L \\
\end{matrix} 
\right)
\label{Eq:Hamiltonian_matrix_H}
\end{eqnarray}
We begin with the definition of winding number $w(h)$ (for $h>0$), given by,
\begin{eqnarray}
w(h) &=& \lim_{L\rightarrow \infty} \frac{1}{2 \pi if} \int_{0}^{2\pi} \partial_\theta \log 
\{ \text{det}[ H(\theta/L,h)- E_b]\} d\theta  \nonumber\\
    &=& \lim_{L\rightarrow \infty}  \int_{0}^{2 \pi} \frac{1}{2 \pi if} (\partial_\theta f) d\theta , 
    \label{Eq:winding-no-def}
\end{eqnarray}
where
\begin{eqnarray}
f(\theta,h)\equiv \text{log}\{\text{det}[H(\theta/L,h)- E_b]\}.
\label{Eq:f_theta_h_def}
\end{eqnarray}
In the above equation, $E_b$ is a base energy, which lies inside the gaps of the energy 
spectrum of the Hermitian AAH Hamiltonian ($h=0)$. From the following discussion, 
it will be clear that the winding number is independent of the choice of $E_b$.

To calculate the winding number, however, it is convenient to use the momentum space form of 
$f(\theta,h) = \rm{log}\{{det}[H_2- E_b]\}$. Here, $H_2$ is connected to $H(\theta/L,h)$
through the similarity transformation $H=R^{-1}H_2R$, where 
$R_{n,l}=(1/\sqrt{L})~\text{exp}(2\pi i\alpha nl)~\text{exp}(-nh+in\theta/L)$. 
The transformed Hamiltonian $H_2$ is given by,
\begin{equation}
H_2=
\left( 
\begin{matrix}
W_1 & \frac{V}{2} & 0 & 0 &  ... & 0 & 0 & \frac{V}{2}  e^{hL}e^{-i\theta}  \\
\frac{V}{2} & W_2 & \frac{V}{2} & 0 &  ... & 0 & 0 & 0 \\
0 & \frac{V}{2} & W_3 & \frac{V}{2} &  ... & 0 & 0 & 0 \\
... & ... & ... & ... & ... & ... & ... & ... \\
0 & 0 & 0 &0 & ... & \frac{V}{2} & W_{L-1} & \frac{V}{2} \\
\frac{V}{2}  e^{-hL}e^{i\theta} & 0 & 0 & 0 & ... & 0 & \frac{V}{2} & W_L \\
\end{matrix} 
\right)\nonumber\\
\end{equation}
where $W_k=2 t' cos(2\pi\alpha k -\eta)$. It is easy to see that in the limit of large $L$ 
($L\rightarrow\infty$), $(H_2)_{L,1} \rightarrow 0$ and $f(\theta,h)$ can be 
written as,
\begin{eqnarray}
f(\theta,h)= \text{log}\Big\{(-1)^{L+1} \Big(V/2\Big)^L e^{hL}e^{-i\theta} +\det(\Theta- E_b)\Big\}, \nonumber \\
\end{eqnarray}
where $\Theta$ is an Hermitian matrix given by, 
\begin{eqnarray}
\Theta=
\left( 
\begin{matrix}
W_1 & \frac{V}{2} & 0 & 0 & 0 & ... & 0 & 0 & 0\\
\frac{V}{2} & W_2 & \frac{V}{2} & 0 & 0 & ... & 0 & 0 & 0 \\
0 & \frac{V}{2} & W_3 & \frac{V}{2} & 0 & ... & 0 & 0 & 0 \\
... & ... & ... & ... & ... & ... & ... & ... & ... \\
0 & 0 & 0 & 0 & 0 & ... & \frac{V}{2} & W_{L-1} & \frac{V}{2} \\
0 & 0 & 0 & 0 & 0 & ... & 0 & \frac{V}{2} & W_L \\
\end{matrix} 
\right)
\label{Eq:Theta-Matrix}
\end{eqnarray}
Since Eq.~(\ref{Eq:Theta-Matrix}) is independent of $\theta$, $f(\theta,h)$ can be written as,
\begin{eqnarray}
\frac{1}{2 \pi if} (\partial_\theta f)  = \frac{1}{2\pi}  \frac{{(-1)}^{L} 
\Big(\frac{V}{2}\Big)^{L} \text e^{hL}e^{-i\theta}}{{(-1)}^{L+1} \Big(\frac{V}{2}\Big)^{L}
\text e^{hL}e^{-i\theta} + \text{det}(\Theta-E_b)}\nonumber\\
\label{Eq:f_partial_theta}
\end{eqnarray}
The value of $\text{det}(\Theta- E_b)$ can be estimated by identifying that 
\begin{eqnarray}
\text{det}(\Theta- E_b)= \text{exp}(g) ,
\end{eqnarray}
where 
\begin{eqnarray}
g=\sum_{L}^{l=1} \text{log} |\lambda_l - E_b|. 
\label{Eq:Def_of_g}
\end{eqnarray}
Here, $\lambda_1,\lambda_2,\cdots,\lambda_L$ are the eigenvalues of the Hermitian matrix $\Theta$ for a 
one-dimensional lattice with $L$ sites. As mentioned earlier, since the base energy is chosen inside the 
small gaps of the energy spectrum of $\Theta$, the value of $g$ never diverges. For a continuous 
energy spectrum, Eq.~(\ref{Eq:Def_of_g}) can be expressed as,
\begin{eqnarray}
g= L \int_{}^{} d\epsilon \rho (\epsilon) \text{log} |\epsilon - E_b |, 
\end{eqnarray}
where $\rho(\epsilon)$ is the density of states. It is now possible to estimate the Lyapunov exponent 
$\gamma$ of an eigenstate of $\Theta$ around 
the neighbourhood of the base energy $E_b$, which is given by \cite{Thouless}, 
\begin{eqnarray}
\gamma = \int d\epsilon \rho (\epsilon) \text{log} |\epsilon - E_b | - \text{log} \Big(\frac{V}{2}\Big).
\end{eqnarray}
Hence, $\text{det}(\Theta- E_b)$ can be written as, 
\begin{eqnarray}
\text{det}(\Theta- E_B)= \Big(\frac{V}{2}\Big)^L \text{exp}(\gamma L). 
\label{Eq:det_Theta_interms_gamma}
\end{eqnarray}
It is well established that for the Hermitian AAH Hamiltonian without the RSO,
$\gamma$ is independent of the base energy and given by \cite{Aubry,Sokoloff},
\begin{eqnarray}
\gamma=\log\Bigg(\frac{2t}{V}\Bigg).
\end{eqnarray}
This follows from the self-duality property of the  Hermitian AAH Hamiltonian without the RSO interaction.
Recently, we have shown that the Hermitian AAH Hamiltonian is 
self-dual in the presence of the RSO \cite{Sahu}, which can also be deduced from Eq.~(\ref{Eq:eigen-eq-quasiparticle}).
Hence, in the presence of the RSO coupling, the Lyapunov exponent $\gamma$ is anticipated 
to be,
\begin{eqnarray}
\gamma=\log\Bigg(\frac{2t'}{V}\Bigg)=\log\Bigg(\frac{2t~\sqrt{1+\Big(\frac{\alpha_y+\alpha_z}{t}\Big)^2}}{V}\Bigg)
\label{Eq:RSO_Lyapunov_Exponent}
\end{eqnarray}
It is now straightforward to see from Eqs.~(\ref{Eq:f_partial_theta}) and (\ref{Eq:det_Theta_interms_gamma}) that,
\begin{eqnarray}
\lim_{L\rightarrow \infty}\frac{1}{2 \pi i f}(\partial_\theta f)= \begin{cases} ~~~~~~0 &\mbox{if } h < \gamma \\
-(1/2\pi) & \mbox{if } h > \gamma \end{cases} 
\end{eqnarray}
From the definition of the winding number $w(h)$ in Eq.~(\ref{Eq:winding-no-def}), we can expect a topological 
transition at a critical value of $h = h_{t}$ ,\\
\begin{eqnarray}
w(h)= \begin{cases} ~~0 &\mbox{if } h < h_{t} \\
-1 & \mbox{if } h > h_{t}. \end{cases} 
\label{Eq:winding_number_final}
\end{eqnarray}

From the analytical result, it is evident that there exists a topological transition in quasicrystals described by 
the $\mathcal{PT}$-symmetric non-Hermitian AAH Hamiltonian with RSO coupling. We have also verified the topological
transition point numerically, and found that indeed there is a topological transition at the crtical value given by $h_t$. 
However, we are going demonstrate that the critical point $h_c$, where the delocalization-localization transition 
takes place, does not coincide with the topological transition point in general, except in some special limiting cases.
Moreover, $h_c < h_t$, whenever they do not conincide. We have also found that the $\mathcal{PT}$-symmetry 
always breaks down at $h_c$.  

We now discuss our main results. Numerical methods to determine the critical point and the winding number
have been discussed in the subsequent sections. The numerical results are for a lattice with $L$=610 sites, and obeying the periodic boundary conditions. In Fig.~\ref{Fig: Phase diagram for h_c}, we have presented the phase 
diagram in the space spanned by the parameters of the Hamiltonian in Eq.~(\ref{Eq:Full_Hamiltonian}). In 
Figs.~\ref{Fig:Phase transition and winding number}(a) and (b), we have presented the results of the numerically obtained critical points 
and compared it with the analytically obtained topological transition point for the two limiting cases: $\alpha_y=0$ and 
$\alpha_z=0$. From the analytical expression obtained in Eq. (\ref{Eq:RSO_Lyapunov_Exponent}), it is obvious that in the 
presence of a single RSO coupling term, the critical value $h_t$ can be expressed for the two different cases as, 
$h_t=\text{log}(2\alpha_z/V)$, and $h_t = \text{log}(2\alpha_y/V)$. 
The topological critical points $h_t$ as a variation of $\alpha_z$ in the former, and with $\alpha_y$ in the latter is 
shown in Figs. \ref{Fig: Phase diagram for h_c}(a) and (b), in which the analytical result matches exactly with the  
critical points $h_c$ computed numerically. However, when the contribution of both the spin-orbit coupling hopping 
amplitudes are non-zero, the topological critical point $h_t$ and the delocalization-localization transition 
point $h_c$ gets separated as can be seen from Figs. \ref{Fig: Phase diagram for h_c}(d) and (e).
Furthermore, it is also quite obvious that on interchanging the non-zero strengths of $\alpha_y$ and $\alpha_z$, 
the existing conditions for the two types of transition remain unaffected. To understand the origin of this 
difference between $h_c$ and $h_t$, we have shown the phase diagram in Figs. \ref{Fig: Phase diagram for h_c}(c) and (f), 
when the tight binding hopping amplitude $t$ is set to zero. It is clear that the critical points $h_c$   
follow exactly the same pattern as Figs. \ref{Fig: Phase diagram for h_c}(a-b). However, the topological critical 
points are remarkably different, suggesting that the difference between $h_c$ and $h_t$ is due to the 
combined effect of the two hopping processes of the RSO Hamiltonian.

To identify the localization of the eigenstates, we use the familiar approach of the Inverse Particpation Ratio(IPR) 
\cite{Mirlin,Wessel} given by,\\
\begin{eqnarray}
	IPR=\frac{\sum_{n} |\psi_n|^4}{(\sum_{n} |\psi_n|^2)^2}
	\label{Eq: IPR}
\end{eqnarray}
It is well established that for extended states, the IPR varies with the system size as $N^{-1} \simeq 0$ 
(in the thermodynamic limit), whereas for localized states, the IPR is system size independent and approaches 1 when 
they are completely localized. For RSO, where the size of the Hamiltonian matrix $N=2L$, these characteristics remain 
unaffected, with an enhancement in the delocalized regime upto the critical point $h_c$. Additionally, in the
$\mathcal{PT}$-symmetric non-Hermitian quasicrystals, we expect the $\mathcal{PT}$-symmetry to be broken at some 
critical value of $h$, when the energy spectrum becomes complex. In 
Fig. \ref{Fig:Phase transition and winding number}(a), the energy spectrum in the complex plane, when a single RSO hopping 
amplitude ($\alpha_z$) is non-zero, has been shown. It is clear that as $h$ crosses beyond the value $0.73 \pm 0.01$, the 
entire energy spectrum becomes complex. For $\alpha_y=0, \alpha_z=0.3$ we estimate this transition point to be 
$h_t=0.73 \pm 0.01$. We have further corroborated our observation by plotting the 
the largest value of $|\text{Im}(E)|$ $vs.$ the complex phase $h$, as shown in in Fig. \ref{Fig:Phase transition and winding number}(b).
The rapid increase in its value from zero suggests the transition from the unbroken PT symmetry  regime to the broken PT 
symmetry  regime at $h_c=0.73 \pm 0.01$. 

To detect the delocalization-localization transition point,  the 
largest and the smallest values of the IPR have been plotted with $h$ in Fig. \ref{Fig:Phase transition and winding number}(c). These results clearly indicate the existence of 
the localization transition exactly at the same critical point $h_c \simeq 0.74$ where the $\mathcal{PT}$-symmetry breaks
down. On substituting $\alpha_y=0$ and $\alpha_z=0.3$ in the analytical expression of the topological critical point 
(Eq.~\ref{Eq:RSO_Lyapunov_Exponent}), we obtain $h_t = 0.736$ (evaluated upto the third decimal place).Interchanging the 
values of $\alpha_y$ and $\alpha_z$ does not alter the critical point and the topological transition point, which are 
identical. Hence, in this case the critical point, the $\mathcal{PT}$-symmetry breaking point and the topological 
transition point are identical. However, this behaviour changes dramatically when both the hopping amplitudes of the 
RSO Hamiltonian are non-zero. 

In the lower panel of Fig.(\ref{Fig:Phase transition and winding number}), we have presented the results when 
both the RSO coupling strengths are considered to be non-zero. Figs. \ref{Fig:Phase transition and winding number}(e-g) show 
the same result corresponding to the Figs. \ref{Fig:Phase transition and winding number}(a-c), except that the localization 
transition point shifts, outstretching the metallic and PT-symmetry unbroken regimes to a higher value of $h_c$. 
In Fig.~\ref{Fig:Phase transition and winding number}(f) and (g), we have also indicated the topological transition point $h_t$ 
anticipated from our analytical result. It is evident that $h_c < h_t$ when both the RSO coupling have non-zero 
amplitudes. In Figs. \ref{Fig:Phase transition and winding number}(d) and (h), we have presented the results of the 
winding number that have been computed numerically. The evidence of the topological transition at the anticipated value of $h$ 
from our analytical result obtained in Eq.~(\ref{Eq:RSO_Lyapunov_Exponent}) is obvious.  

\indent \emph{Conclusions-} In summary, we have demonstrated that the delocalization-localization and the PT 
symmetric breaking transitions are not connected to a topological transition in the $\mathcal{PT}$-symmetric non-Hermitian
quasicrystals. The topological transition point $h_t$ has been obtained analytically by mapping the Hamiltonian in the presence 
of RSO to the original $\mathcal{PT}$-symmetric non-Hermitian AAH Hamiltonian and have been verified numerically. 
The critical point $h_c$ has been determined numerically, and we have found that in general the delocalization-localization 
transition and the PT symmetric breaking takes place earlier than the topological phase transition, except in some special cases,
although the delocalization-localization transition and the PT symmetric breaks simultaneously.
\indent\emph{Acknowledgement}- This work is supported by SERB (DST), India 
(Grant No. EMR/2015/001227). A. C would like to thank CSIR(HRDG) for providing financial assistance 
(File No.-09/983(0047)/2020-EMR-I).
\bibliography{reference} 
\end{document}